\documentclass[twocolumn,prl,showpacs,superscriptaddress,floatfix]{revtex4}

\usepackage{amsmath}
\usepackage{verbatim}
\usepackage{graphicx}

\newcommand{\ket}[1]{\ensuremath{|#1\rangle}}
\newcommand{\bra}[1]{\ensuremath{\langle#1|}}
\newcommand{\nuc}[2]{\mbox{${}^{#1}\rm #2$}}
\newcommand{\units}[2]{\mbox{$#1\,\text{#2}$}}
\newcommand{\half}{\mbox{$\frac{1}{2}$}}
\newcommand{\degree}[1]{\ensuremath{#1^\circ}}
\newcommand{\para}{\textit{para}}
\newcommand{\Para}{\textit{Para}}
\newcommand{\pHH}{\para-hydrogen}
\newcommand{\ortho}{\textit{ortho}}
\newcommand{\HH}{\mbox{$\text{H}_2$}}
\newcommand{\hydride}{\mbox{$\text{Ru(H)}_2\text{(CO)}_2\text{(dppe)}$}}
\newcommand{\precursor}{\mbox{$\text{Ru(CO)}_3\text{(dppe)}$}}
\newcommand{\intermediate}{\mbox{$\text{Ru(CO)}_2\text{(dppe)}$}}
\newcommand{\dn}{\downarrow}
\newcommand{\up}{\uparrow}

\begin{document}
\title{Preparing high purity initial states for\\ nuclear magnetic resonance
quantum computing}
\author{M.~S.~Anwar}
\affiliation{Centre for Quantum Computation, Clarendon Laboratory,
University of Oxford, Parks Road, OX1 3PU, United Kingdom}
\author{D.~Blazina}
\affiliation{Department of Chemistry, University of York,
Heslington, York, YO10 5DD, United Kingdom}
\author{H.~A.~Carteret}\email{cartereh@iro.umontreal.ca}
\affiliation{Institute for Quantum Computing and Department of
Combinatorics and Optimization, University of Waterloo, Waterloo,
Ontario, N2L 3G1, Canada} \affiliation{Physics Department,
Imperial College, Prince Consort Road, London, SW7 2BZ, United
Kingdom}
\author{S.~B.~Duckett}\email{sbd3@york.ac.uk}
\affiliation{Department of Chemistry, University of York,
Heslington, York, YO10 5DD, United Kingdom}
\author{T.~K.~Halstead}
\affiliation{Department of Chemistry, University of York,
Heslington, York, YO10 5DD, United Kingdom}
\author{J.~A.~Jones}\email{jonathan.jones@qubit.org}
\affiliation{Centre for Quantum Computation, Clarendon Laboratory,
University of Oxford, Parks Road, OX1 3PU, United Kingdom}
\author{C.~M.~Kozak}
\affiliation{Department of Chemistry, University of York,
Heslington, York, YO10 5DD, United Kingdom}
\author{R.~J.~K.~Taylor}
\affiliation{Department of Chemistry, University of York,
Heslington, York, YO10 5DD, United Kingdom}
\date{\today}
\begin{abstract}
Here we demonstrate how \pHH\ can be used to prepare a two-spin
system in an almost pure state which is suitable for implementing
nuclear magnetic resonance (NMR) quantum computation.  A
\units{12}{ns} laser pulse is used to initiate a chemical reaction
involving pure \pHH\ (the nuclear spin singlet of \HH). The
product, formed on the $\mu\text{s}$ timescale, contains a
hydrogen-derived two-spin system with an effective spin-state
purity of 0.916. To achieve a comparable result by direct cooling
would require an unmanageable (in the liquid state) temperature of
\units{6.4}{mK} or an impractical magnetic field of
\units{0.45}{MT} at room temperature. The resulting spin state has
an entanglement of formation of 0.822 and cannot be described by
local hidden variable models.

\end{abstract}
\maketitle

\paragraph{Introduction.}
While quantum computing \cite{BDiV1} offers the
potential of using new quantum algorithms to tackle problems that
are intractable for classical processors, its implementation
requires the development of quantum devices, which are as yet
unavailable. The most complex implementations of quantum
algorithms to date have used techniques adapted from nuclear
magnetic resonance (NMR) spectroscopy \cite{Cory, Gershenfeld,
JonesRev, Vandersypen}, but current liquid state NMR approaches
cannot be extended to systems with many quantum bits, as it is not
possible to prepare pure initial states by directly cooling the
spin system into its ground state \cite{Warren}.  Furthermore, it
has been shown that current NMR experiments involve only separable
states \cite{Braunstein}, and thus could in principle be described
by local hidden variable models.

The conventional approach in NMR quantum computing \cite{JonesRev} is
to use an ensemble of spins, and to prepare a pseudo-pure ground state
\cite{Cory, JonesRev} of the form
\begin{equation}
\rho=(1-\varepsilon)\frac{\openone}{2^n}+\varepsilon\ket{0}\bra{0}
\end{equation}
where $\openone/2^n$ is the maximally mixed state of an $n$-spin
system, and $\varepsilon$ is the polarization of the state.  In
the high temperature regime this approach is exponentially
inefficient \cite{Warren}.  Furthermore if the polarization lies
at or below a critical bound then any apparently entangled states
prepared from the pseudo-pure state are in fact separable
\cite{Braunstein}; for two qubits \cite{Peres, H3sep, Horodecki2}
this bound is $\varepsilon=1/3$, corresponding to a fractional
population of $1/2$ in the ground state and $1/6$ in each of the
three other eigenstates. A radically different approach is to
prepare initial states using non-thermal means \cite{NMRcrit},
\textit{e.g.,} by using the pure singlet nuclear spin state isomer
of \HH\ (called ``\pHH '') \cite{Bowers,Natterer, ApppH2,
DuckBlaz} as a cold spin-state reservoir.

\paragraph{\textit{Para}-hydrogen induced polarization.}
A pure singlet nuclear spin state can be described
using product operator notation \cite{Sorensen} as
\begin{equation}\label{singlet}
\half(\half E -2I_xS_x-2I_yS_y-2I_zS_z) \\
\end{equation}
where $E=\openone\otimes\openone, \;
I_x=\half(\sigma_x\otimes\openone), \;
S_x=\half(\openone\otimes\sigma_x)$ and so on. The existence of
\pHH\ is a consequence of the Pauli principle \cite{Shankar},
which requires the overall wave function of the molecule to be
antisymmetric with respect to particle interchange. Dihydrogen
molecules in even rotational states ($\text{J}=0,\;2,\;\ldots$)
possess an antisymmetric nuclear wave function and correspond to
nuclear spin singlets
($S_0=(\mbox{\ket{\!\up\dn}}-\mbox{\ket{\!\dn\up}})/\sqrt{2}=\Psi^-$,
termed \para). Molecules in odd rotational states
($\text{J}=1,\;3,\;\ldots$) have symmetric nuclear wave functions
and consist of the three nuclear spin triplets
($T_0=(\mbox{\ket{\!\up\dn}}+\mbox{\ket{\!\dn\up}})/\sqrt{2}=\Psi^+$,
$T_{-1}=\mbox{\ket{\!\up\up}}$, and
$T_{+1}=\mbox{\ket{\!\dn\dn}}$, termed \ortho). Note that the
$T_{\pm1}$ states are \textit{not} the same as
$\Phi^\pm=(\mbox{\ket{\!\up\up}}\pm\mbox{\ket{\!\dn\dn}})/\sqrt{2}$
(the other two Bell states), but that an equal mixture of $T_{+1}$
and $T_{-1}$ is also an equal mixture of $\Phi^+$ and $\Phi^-$.

Isolation of the \pHH\ spin isomer is possible because spin-isomer
interconversion is forbidden by angular momentum selection rules.
Adsorption onto a suitable surface breaks the symmetry of the \HH\
molecules, allowing spin-isomer interchange. The new \ortho/\para\
ratio therefore \textit{remembers} the temperature of the last
conversion surface encountered. Upon moving away from this surface
interconversion is again suppressed. A temperature of
\units{20}{K} is sufficiently low to form the $\text{J}=0$ state
and hence produce pure \pHH. The resulting
\para-\HH\ molecule cannot be used directly, as it is NMR silent
due to its high symmetry. By means of a chemical reaction,
producing a new molecule, the two hydrogen atoms can be made
distinct (I and S) and \textit{can} be separately addressed. This
phenomenon is well known from mechanistic NMR studies of catalytic
hydrogenation and hydroformylation, where it is usually referred
to as \pHH\ induced polarization, or PHIP \cite{Bowers, Natterer,
ApppH2, DuckBlaz}.

In many previously described PHIP experiments the addition process
is slow in comparison with the frequency difference between the I
and S spins of the reaction product.  This causes the off-diagonal
terms in the density matrix to dephase, resulting in an equal
mixture of singlet and $T_0$ triplet states in the spin ensemble
\cite{dephaseXY}, which is separable \cite{Peres, H3sep,
Horodecki2}.  One possible method that has been suggested to
overcome this problem is to perform the slow addition while
applying an isotropic mixing sequence \cite{MLEV} to remove the
dephasing effect \cite{Bargon}.  In principle (and neglecting
relaxation effects) this should completely conserve the hydrogen
spin state, and this method has been used \cite{Bargon} to achieve
states with $\epsilon\approx 0.1$.

A much simpler approach is to ensure that addition is rapid in
comparison with the dephasing and relaxation timescales. This
requires addition to a highly reactive species, but a fast
reaction would go to completion and the spin system would decohere
before the sample could be studied. In this paper we solve this
problem by using photochemistry to prepare a reagent which will
react instantaneously with \pHH\ to produce the molecule for
examination.  This means we can start and stop the formation of
the reactive species in a controlled manner. We then show that the
resulting two-spin system forms in an almost pure state.

\begin{figure*}
\includegraphics[scale=0.8]{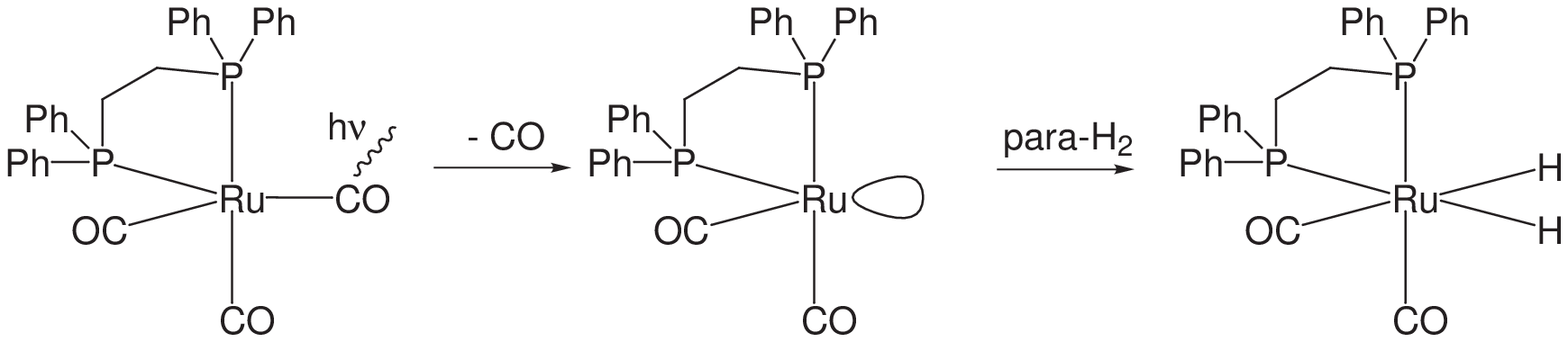}
\caption{The reaction scheme used to generate \hydride\ with an
almost pure initial spin state.  A UV photon knocks one carbonyl
group off the \precursor\ precursor to generate an unstable
intermediate, which immediately adds hydrogen to give the desired
product. Since
\para-\HH\ has a pure singlet initial state, and addition occurs
with retention of spin state, the product should also have a pure
singlet spin state.}
\end{figure*}

\paragraph{Experimental methods.}
In this study the precursor molecule was \precursor, where dppe
indicates 1,2-bis(diphenylphosphino)ethane. This was prepared from
$\text{Ru}_3(\text{CO})_{12}$ by warming a benzene solution to
\units{373}{K} under 30 atmospheres of CO in the presence of three
equivalents of dppe \cite{Schott}. The \precursor\ was dissolved
in $\text{d}_6$-benzene in a \units{5}{mm} NMR tube fitted with a
Young's valve to permit attachment to a vacuum line.  Dissolved
gases were removed by freeze-pump-thaw cycles and the tube was
covered in foil to exclude light. \Para-hydrogen was prepared at a
temperature of \units{20}{K} using a charcoal-based \ortho-\para\
interconversion catalyst; at this temperature the thermal state is
essentially pure \para. The \pHH\ was then introduced to the
sample; as the catalyst is no longer present \ortho-\para\
interconversion is suppressed. After warming, shaking ensures that
the \HH\ gas (with a pressure of about 3 atmospheres) dissolves.

The NMR tube was placed in a \units{400}{MHz} NMR spectrometer
fitted with a \nuc{1}{H}/\nuc{31}{P} tuned NMR probe equipped for
\textit{in situ} photolysis \cite{Godard}. (The transfer was
performed in a darkened room to prevent premature photolysis by
ambient light.) The reaction was initiated by a \units{12}{ns}
pulse of \units{308}{nm} UV light (pulse energy \units{32}{mJ})
from an MPB Technologies MSX-250 pulsed XeCl excimer laser
triggered by the NMR spectrometer.  This generates the reactive
intermediate \intermediate, \textit{in situ} from its stable
precursor \precursor\ by laser flash photolysis (see Fig.~1).
Subsequent reaction of \intermediate\ with \HH\ occurs on the
sub-microsecond timescale \cite{Cronin} and leads to the product
of interest, \hydride.

The two hydride resonances of the product appear at
$\delta=-7.55\,\text{ppm}$ (spin I) and $\delta=-6.32\,\text{ppm}$
(spin S). For the analysis a spin I selective pulse was
implemented using two hard \degree{90} pulses separated by a delay
of $1/(4\delta\nu)$, where $\delta\nu=492\,\text{Hz}$ is the
difference between the resonance frequencies, with relative phases
of \degree{135} and with the RF frequency centered on the midpoint
of the two resonances. Such pulses, based on Jump and Return
sequences, have been described previously \cite{approxcount}. The
GARP sequence \cite{GARP} was applied throughout signal
acquisition to remove couplings to \nuc{31}{P} nuclei. To avoid
the necessity for complete quantum state tomography \cite{Chuang}
a filtration sequence was developed which has no effect on the
desired singlet state, but dephases most other states.

While the signal produced by the \pHH\ is easily seen in one scan,
the thermal signal is extremely weak and so difficult to measure
directly.  It was therefore necessary to increase its intensity by
increasing the amount of \hydride\ in the sample. This was
achieved by applying a further 999 laser pulses to produce more
\hydride. Even then we needed 3072 scans for the calibration
spectrum; these were separated by an interval of \units{20}{s},
which is much greater than five times the measured $\text{T}_1$ of
\units{1.7}{s} and so saturation effects can be ignored. The
spectra were processed by homewritten software and analysed by
integration.  The \pHH\ spectrum comprises a pair of antiphase
doublets, which partially cancel \cite{remsplitt}, and so direct
integration will result in an underestimate of the signal
intensity; to reduce such effects the spectrum was J-doubled
\cite{McIntyre} four times before integration. Even after
J-doubling and integration slight imbalances were visible between
the two multiplets, and these imbalances can be analysed to
determine the imbalance between $T_0$ and $T_{\pm1}$ triplet
states.

\paragraph{Results.}
Laser photolysis of the \precursor\ leads to the product \hydride.
If the two hydrogen nuclei have inherited the nuclear singlet from
the \pHH, then a selective \degree{90} $I_y$ pulse will yield the
observable NMR terms
\begin{equation}\label{xzterms}
\half(-2I_xS_z+2I_zS_x).
\end{equation}
This corresponds to a pair of antiphase doublets, with intensities
$\pm1/2$.  This pattern is indeed seen (Fig.\ 2), but to show that
we have an essentially pure singlet state it is necessary to
determine the \textit{intensity} of the signal as well as its
form. We calibrated our signal against a standard provided by the
thermal state of the same spin system, obtained by allowing the
spin system to relax.  At high temperatures the thermal state is
\begin{equation}
\half(\half E+\half B[I_z+S_z])
\end{equation}
where $B=h\nu/kT$ and $\nu$ is the Larmor frequency of the spins.
After a \degree{90} pulse the NMR observable terms are
\begin{equation}
\mbox{$\frac{1}{4}$}B(I_x+S_x),
\end{equation}
that is a pair of inphase doublets with intensity $B/4$, so that a
pure state will give a signal $2/B$ times larger than a thermal
state. For our system $\nu=\units{400}{MHz}$ and
$T=\units{295}{K}$, and so the \pHH\ signals could be up to 31028
times more intense than the thermal signals.
\begin{figure}
\includegraphics[scale=1.0]{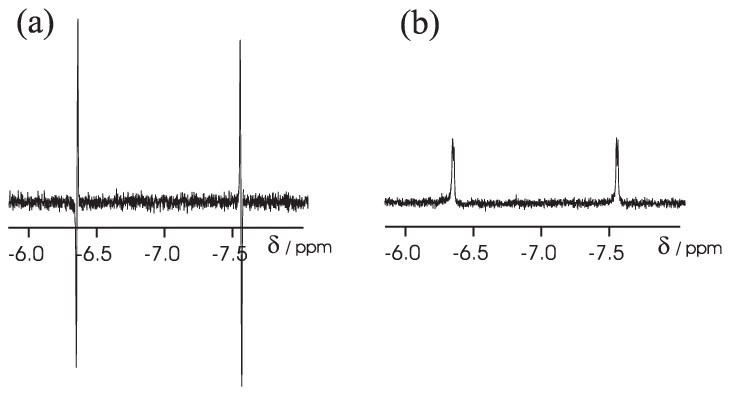}
\caption{The \pHH\ enhanced and calibration spectra of \hydride.
The \pHH\ spectrum (a) is a single scan after a single laser
flash, while the calibration spectrum (b) is the sum of 3072 scans
after 1000 laser flashes. The calibration spectrum has been
divided by $3072\times1000$ and then multiplied by the theoretical
maximum enhancement (31028, see main text) before plotting, so
that the two spectra should show the same intensity.  In fact the
\pHH\ spectrum (a) is even more intense than na{\"\i}vely
predicted.}
\end{figure}

In order to examine the form of the signal, we used a filtration
sequence (comprising two periods of length $1/\delta\nu$, under a
field gradient separated by a hard \degree{90} pulse) which is
closely related to ``twirl'' operations \cite{EntForm,
Purification}.  After this, the density matrix is Bell-diagonal,
comprising a mixture of singlet and triplet states, with equal
amounts of the parallel-spin triplet states $T_{+1}$ and $T_{-1}$.
Since this sequence had little effect on the spectrum, the initial
density matrix had a similar form.

Our results (Fig. 2) show an apparent enhancement of about 77000,
significantly higher than expected.  This discrepancy arises
because the NMR probe is not sensitive to the entire sample,
but only to that within the RF coil.  The hydride forms within the
coil region because of the position of the mirror that introduces
the UV light, as confirmed by one-dimensional NMR imaging, but is
then distributed throughout the sample by convection and
diffusion, so that in the calibration spectrum only a fraction of
the hydride is detectable. As expected the directly measured
enhancement (that is, before correction) shows a linear dependence
on the total sample volume (data not shown). The active volume
fraction can be estimated using geometrical arguments based on the
relative length of the NMR sample and the RF coil.  The actual
value depends on the length of the NMR sample, but was 0.368 in
the experiment described.  After correcting for this active volume
fraction the observed enhancement is consistent with a
polarization of $\varepsilon=0.916\pm0.019$.

\paragraph{Relaxation.}
The measured relaxation and decoherence times of the \nuc{1}{H}
nuclei in the \pHH\ state ($\text{T}_1=\units{1.7}{s}$,
$\text{T}_2=\units{0.58}{s}$) are indistinguishable from those in
the thermal state, indicating that the high polarisation does not
affect the relaxation properties of the molecule.  No effects of
radiation damping were observed, reflecting the extremely low
concentration of the hydride.

\paragraph{Entanglement.}
As well as the polarization enhancement, the use of PHIP
initializes the system directly into an entangled state. The
entanglement threshold of $1/3$ is valid only for mixtures of a
Bell state with the maximally mixed state (Werner states).  Our
density matrix could contain arbitrary states mixed with the
singlet and the threshold depends on what is mixed in. The states
that most effectively destroy the entanglement of the singlet are
$T_0$ and equal mixtures of $T_{+1}$ and $T_{-1}$; these are also
the only states which survive the filtration sequence.

The positivity of the partial transpose test \cite{Peres} is a
necessary and sufficient condition for separability for two qubit
states \cite{H3sep}.  It can be shown \cite{Horodecki2} that for a
system comprising a mixture of the singlet and some convex
combination of $T_0$ and an equal mixture of $T_{+1}$ and $T_{-1}$
the state will always be entangled if the total amount of singlet
exceeds $1/2$.
Detailed analysis of our data shows that the density matrix is not
quite a Werner state: instead the state comprises 93.7\% $S_0$,
4.5\% $T_0$, 0.9\% $T_{+1}$ and 0.9\% $T_{-1}$.  The excess
population of $T_0$ suggests the presence of some phase
decoherence process that we have yet to identify, but this excess
has no effect on the entanglement of formation \cite{EntForm,
concurrence}, which would be 1 for a pure singlet and for our
system is $0.822\pm0.039$.

Our results demonstrate that liquid-phase NMR can produce
entanglement. The question of whether entanglement really is a
necessary resource for universal quantum computers can only be
answered by mathematical proofs of their scaling characteristics
with and without entanglement. These can only be defined
rigorously in the limit as the size of the problem instance (and
hence the computer solving it) tends to infinity.  Any actual
experiment can only be performed on a finite-sized system, and can
therefore neither prove nor disprove the correctness of the
arguments in \cite{Braunstein}. We can only show that this
particular objection can no longer be levelled at liquid-phase NMR
quantum computation.

\paragraph{Conclusions and further work.}
We have shown that \pHH\ can greatly benefit NMR quantum computing
by the generation of almost pure ($\varepsilon=0.916\pm0.019$)
initial states on demand, without the need for lengthy preparation
sequences. To achieve a comparable result by direct cooling would
require an unmanageable (in the liquid state) temperature of
\units{6.4}{mK} or an impractical magnetic field of
\units{0.45}{MT} at room temperature.

Conventional liquid-phase NMR has poor scaling characteristics at
low polarization \cite{Warren}, but our effective spin temperature
of less than \units{6.4}{mK} is much less than the \units{1}{K}
threshold suggested by Warren for efficient initialization. This
approach can in principle be scaled up by adding $M$ molecules of
\pHH\ to a single precursor, effectively synthesizing a quantum
computer with $2M$ qubits in a pure initial state, and we are
currently seeking to implement these ideas.  We also expect that
the ability to rapidly generate highly spin-polarised hydride
species on demand will prove useful in other \pHH\ based studies
of reaction mechanisms.

\begin{acknowledgments}
We thank the EPSRC for financial support.  MSA thanks the Rhodes
Trust for a Rhodes Scholarship.  HAC thanks MITACS, The Fields
Institute and the Canadian NSERC CRO for financial support.  We
are also grateful for helpful discussions with Prof. R. N. Perutz
and Prof. K. Muller-Dethlefs, and experimental advice from P. L.
Callaghan and K. A. M. Ampt.
\end{acknowledgments}

\end{document}